# ZERO ENERGY TRAVEL

*O. Ahmad[1], A. Kiring[2], A. Chekima[3]*
[1,2,3]Universiti Malaysia Sabah, 88400 Kota Kinabalu
othmana@gmail.com

**ABSTRACT**

It is fundamentally possible to travel with zero energy based on Newton's Laws of Motion. According to the first law of motion, a body will continue to travel for infinite distance unless it is acted upon by another force. For a body in motion, the force which stops perpetual motion is friction. However, there are many circumstances that friction is zero, for example in space, where there is vacuum. On earth, gravity makes objects to be in constant contact with each other generating friction but technology exists to separate them in the air using powerful magnetic forces. At low speeds, the friction caused by air is minimal but we can create vacuum even on land for high speed travel. Another condition for travelling is for it to stop at its destination. On land, we can recover the kinetic energy back into electrical energy using brushless permanent magnet generators. These generators can also convert electric energy into kinetic energy in order to provide motion. This article reviews technologies that will allow us to travel with zero energy. It is easier to do it on land but in the air, it is not obvious.

*Keywords*: electrical energy, Faraday's law, gravity, kinetic energy, Newton's laws, potential energy, zero energy,

## INTRODUCTION

IN this day of dwindling natural resources, it is vital that we conserve energy. Theoretically, it requires zero energy to move from one place to another.

Stored energy is converted to kinetic energy at the start of a journey. This kinetic energy is what moves the object. When we want to stop at our destination, we simply put the kinetic energy back to the stored energy.

From the principle of the conservation of energy:

Stored Energy1 = Kinetic Energy

Kinetic energy, $E_k$ is given by:

$$E_k = \tfrac{1}{2} mv^2 \qquad (1)$$

Where $m$ is the mass of the object and $v$ is the velocity of the object.

Equation (1) implies that, if we want to move faster, we need to apply more energy.

If we want to stop, we must convert $E_k$ to another form of stored energy, called Stored Energy2. Then the velocity of the object is zero but the energy is transferred to another form. If we do not store the energy, it will dissipate as other forms of energy such as heat.

Kinetic Energy = Stored Energy2

Without any loss, Stored Energy1 = Stored Energy2, i.e. we manage to recover our original energy in the form that we can make use of later on.

Fig. 1 shows the energy losses in a typical car derived from [1].

Most of the losses are due to energy conversion from chemical to mechanical, via heat. Losses include sound and heat as well as friction. For land based and almost flat travelling, there is only the electric to kinetic energy conversion in electric cars.

For lifting devices, energy recovery is possible when the object is going down. The motor behaves like a generator that provides a reverse force to stop the lift from crashing to the ground.

In an elevator that has a counter weight, there is no need to overcome the gravitational force of the elevator body. The lifting motor only needs to provide extra force to push additional weight up [2]. "The most efficient equipment(elevator) uses regenerative braking to feed electric energy back into the building instead of dissipating as heat", Quoted from [3].

Extra energy, $E_l$ to lift mass $m$ is:

$E_l = g \times m$, where $g$ is the gravitational constant.

The energy recovered during the downward force should also be equal to $E_l$.

For flying machines, the only way to recover kintetic energy, is by using propellers that behave like wind turbines in wind electrical power generators when a plane lands. There is no known way to recover the energy generated by jet engines. The nearby future of zero energy travel is only for ground based machines or propeller driven aeroplanes.

However, if it is possible to lift objects with zero energy, it should be possible to lift space ships into space and then launch the space ships in the vacuum environment. The ships can be accelerated magnetically. On arrival, the ships decelerate while absorbing energy in space using loops of various sizes so as to avoid collisions at high cruising speeds.

Alternatively, the elevator can be in a form of a linear accelerator that will launch the ship into space. It will also receive the space ship on arrival, recovering the energy.

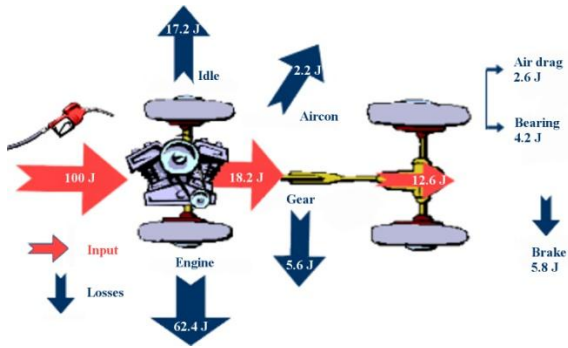

**Fig. 1** Energy losses in a typical car in a typical driving cycle per 100 J of input energy.

## LITERATURE REVIEW

### Review Stage

It is surprising that there is very little information about zero energy travel. Searching through the web using Google, will show references such as [4] that refers to an article in [5] and [6], that touts the Casimir effect as a possible way to travel with zero energy cost. There is no luck with Yahoo, Bing, Google scholar, Science direct and Scopus searches. The most that they can offer are articles for interstellar travel that are still experimental. It is surprising when brushless motors/generators have the potential to achieve zero friction and zero conversion loss. In fact, brushless motors/generators have already achieved very low energy losses already [7].

These brushless motors achieve their high efficiency level due to the magnetic field effect. The magnet isolates the metal parts from each other. Mechanical friction is caused by contacts between metals. Brushless motors avoids frictional contact with any metal when running. Their components are separated by a magnetic field similar to the magnetic levitation in trains. [8] describes the latest Maglev trains and the Maglev trains installed in China. There are even superconducting coils used in the Japanese Maglev.

Based on the data on the Maglev, there is little energy saving in these Maglev despite having such low friction and thus lower losses in heat and sound. There is another condition for zero energy travel. The energy must be recoverable. Kinetic energy recovery can be done by electric motors that fundamentally function as generators at the same.

### Basic equations

**Production of Force on a current carrying conductor,**
$F = i(l \times B)$
where:

$i$ – represents the current flow in the conductor

$l$ – length of wire, with direction of $l$ defined to be in the direction of current flow

$B$ – magnetic field density

**Voltage induced on a current carrying conductor moving in a magnetic field** (Faraday's law of induction),
$e_{ind} = (v \times B) \, l$

where:

$v$ – velocity of the wire relative to magnetic flux

$B$ – magnetic field density

$l$ – length of the wire in the magnetic field

NOTE:

The same machine acts as both motor and generator. The only difference is whether the externally applied force is in the direction of motion (generator) or opposite to the direction of motion (motor). Refer to textbooks such as [9].

### Lossless

These equations imply lossless conversions. Practical motors and generators have losses but these are due to electrical resistance, friction, heat and sound losses. Electrical energy is converted to all these other forms of energy. Brushless motors have no friction in their operating conditions, simply because there is no contact between materials. There is still air but this can be eliminated or ignored as its value is too small at the operating region compared to other losses such as air friction of the body of the vehicle or motor itself. The air drag can be eliminated altogether if we operate in a vacuum.

This vacuum can be created in tunnels where vehicles can move at extremely high speed. We can make the vehicle so long that it occupies all space in the tunnel thus making it easier to create a vacuum. We can use airlocks to reduce air from the tunnel slowly while using the vehicle itself as a plunger to remove some of the air as well. These are just suggestions as to what can be done to reduce friction to the barest minimum. Although we cannot reduce friction to absolute zero, if we reduce it enough, it will certainly save lots of energy.

Another potential loss is electrical resistance. We can remove electrical resistance by using superconductors. The Siemens superconducting generators aims to improve generator efficiency by 0.5 percentage points to 99.5 percent [10]. A patent had been issued for a superconducting wind turbine generator [11].

The bottleneck could be the energy recovery. Battery storage efficiency is not very high but there are alternative electrical storage systems that can store at close to 100% efficiency. A promising system is the ultracapacitor [12]. It can be used at the expense of energy storage capacity [13]. We just need more of these devices. Its mass is more but mass is not a factor in zero energy travel. More mass only requires more energy to move at the same speed as a vehicle with less weight.

To maintain vacuum and low temperature environments requires initial energy input. This initial input can be considered as a fixed investment towards travelling. Leakages in insulation and sealing, will need maintenance energy, in addition to equipment servicing.

## OBSERVATIONS AND DISCUSSION

For a conventional mechanical or electrical car, there is no energy recovery so the loss is 100%. Toyota Prius has some form of energy recovery. The Prius brake regeneration efficiency is assumed to be 41% as indicated by NREL's ADVISOR simulation [14]. Table 2 makes an assumption that regenerative braking efficiency is 99.5%, similar to the proposed Siemens superconducting electric generator [10].

TABLE 1

| Forces are measured in Newton, N, when there is a load of 10000 N, corresponding to approximately 1000 kg weight. | Coefficient of friction | Startup Estimate | Load(N) | Moving Frictional Force | Startup Frictional Force | Reference |
|---|---|---|---|---|---|---|
| Rolling bearing(lowest) | 0.001 | 0.002 | 10000 | 10 | 20 | [15] |
| Rolling bearing(lowest) | 0.005 | 0.01 | 10000 | 50 | 100 | [15] |
| Steel on steel | 0.57 | 0.74 | 10000 | 5700 | 7400 | [16] |
| ice on ice | 0.03 | 0.1 | 10000 | 300 | 1000 | [16] |
| teflon on teflon | 0.04 | 0.04 | 10000 | 400 | 400 | [16] |
| Synovial joints in humans | 0.003 | 0.01 | 10000 | 30 | 100 | [16] |
| AlMgB14 | 0.02 | 0.02 | 10000 | 200 | 200 | [17] |
| Brushless motor | 0.00001 | 0.74 | 10000 | 1 | 7400 | Estimated to be 1/100 of the minimum. |

TABLE 2

| | Urban | Narrow | Vacuum | 60MPH | Units |
|---|---|---|---|---|---|
| Total energy demanded at wheels: | 178,654 | 168,920 | 422,411 | 2,909,770 | J |
| Consumed by Aerodynamic Drag: | 23,867 | 11,933 | 25,472 | 2,547,200 | J |
| Consumed by Rolling Resistance: | 107 | 107 | 1,635 | 1,635 | J |
| Consumed by Braking: | 154,680 | 156,879 | 395,303 | 360,934 | J |
| Distance traveled: | 0.6 | 0.6 | 9.4 | 9.4 | miles |
| Net fuel economy (engine 80%, drivetrain 99.5%): | 331 | 350 | 2,136 | 310 | mi/gal |
| With regenerative braking at 99.5 percent: | 2,391 | 4,614 | 31,021 | 354 | mi/gal |
| Curb Weight: 1000 | | | | | kg |
| Payload: 100 | | | | | kg |
| Drag Coefficient: 0.2 | | | | | |
| Frontal Area: | 2.0 | 1.0 | 2.0 | 2.0 | m² |
| Rolling Resistance Coefficient: 0.00001 | | | | | |
| Wheel Diameter: 0.75 | | | | | m |
| Rotation Factor: 1 | | | | | |
| Air Density: kg/m3 | 1.2 | 1.2 | 0.012 | 1.2 | |
| Fuel heating value: 114132 | | | | | btu/gal |
| Driving Cycle: [18] | ECE-ELEM | ECE-ELEM | 60MPH | 60MPH | |
| Source: [19] | http://www.virtual-car.org/wheels/wheels-road-load-calculation.html | | | | |
| Date: | 26/4/2011 | | | | |

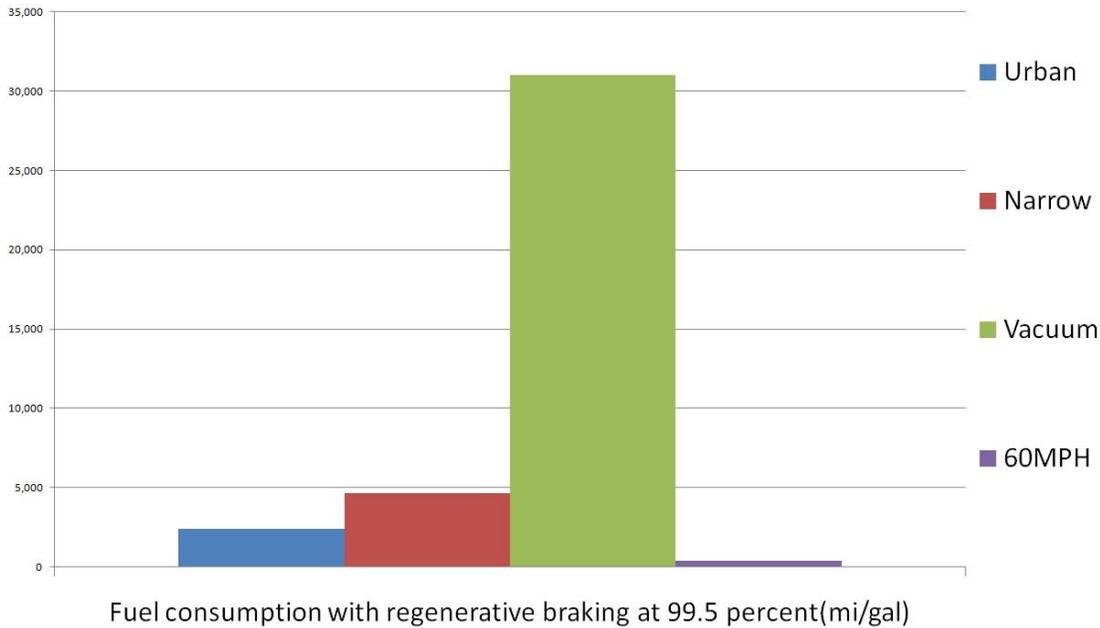

Fuel consumption with regenerative braking at 99.5 percent(mi/gal)

**Fig. 2  A graphical form of selected items from Table 2.**

Energy recovery can only be achieved when braking. Table 2 only shows the air drag, rolling friction and braking energies. Braking energies constitute the useful kinetic energy and can be recovered. Rolling resistance is calculated assuming a friction coefficient that is 1/100th of a roller bearing although there is no friction in a magnetically levitated bearing in a brushless motor (Table 1). The loss can be due to air drag that surrounds the wheel. Air drag and rolling resistance energies cannot be recovered so is deemed as lost.

Air drag is low at low speeds so in urban driving cycle, there is less concern for it. At the 60MPH driving cycle, air drag has become a large contribution of the total energy usage and this energy cannot be recovered. In a vacuum, assumed to be 1/100th of atmospheric pressure, air drag loss is much lower as shown in Fig. 2. At a normal atmospheric pressure, air drag loss is so much higher than braking that regenerative braking does not improve mileage per gallon much.

Mileage per gallon is used because the source of these calculations use these units. Petrol consumption level is included in order to give an indication on the fuel saving. The assumption is based on the use of a hybrid power plant that can achieve up to 80% combined combustion efficiency. This can only be achieved if batteries with a conversion efficiency of 90% and higher are used.

For a loss free environment, highly efficient battery, backed up by highly efficient capacitors can achieve a much higher efficiency but is harder to be used as an example to indicate fuel economy. A petrol consumption equivalent is shown in the fuel economy rows in units of miles / gallon (0.43 km / litre) for a hybrid engine. A fuel economy figure of 31 thousand miles per gallon is achievable in a vacuum.

In space, the quality of vacuum is much higher so there is even less air drag. Lower losses at higher speeds are possible high up above ground. Magnetic principles can also be used to propel spaceships and later recover back the energy used when stopping the spaceships.

CONCLUSIONS

According to Newton's laws of motion, it is possible to travel with zero energy. The conditions are that there is zero friction and the kinetic energy can be recovered. However Faraday's law of induction shows that this energy can be recovered without any loss provided that there is zero loss due to electrical resistance.

Permanent magnet brushless motors can operate without any metal to metal contact so there is no friction. Superconducting materials allows electrical currents to flow with zero electrical resistance. Air drag will be zero in a vacuum. Therefore it is not impossible to achieve zero energy travel using available techniques.

Of course we have not reached perfection yet and it will be very difficult and time consuming but even if we can achieve a level of efficiency of 99.5%, the savings will be immense, equivalent to thousands of miles per gallon of petrol. It will certainly help in conserving our environment and we don't need exotic physics to do it.

APPENDIX

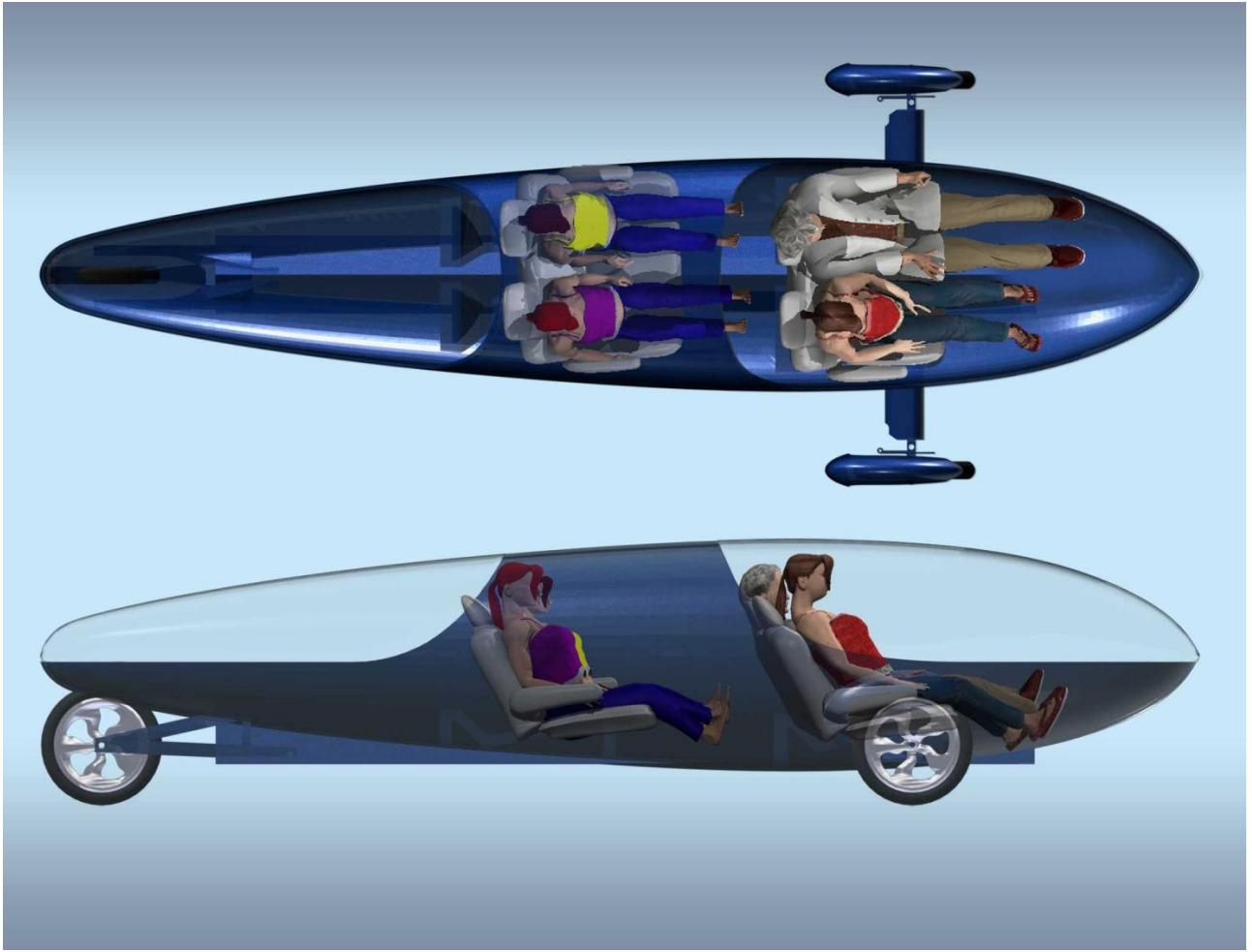

**Fig. 3  Tritrack street. A possible future car. (http://www.electric-bikes.com/betterbikes/tritrack.html)**

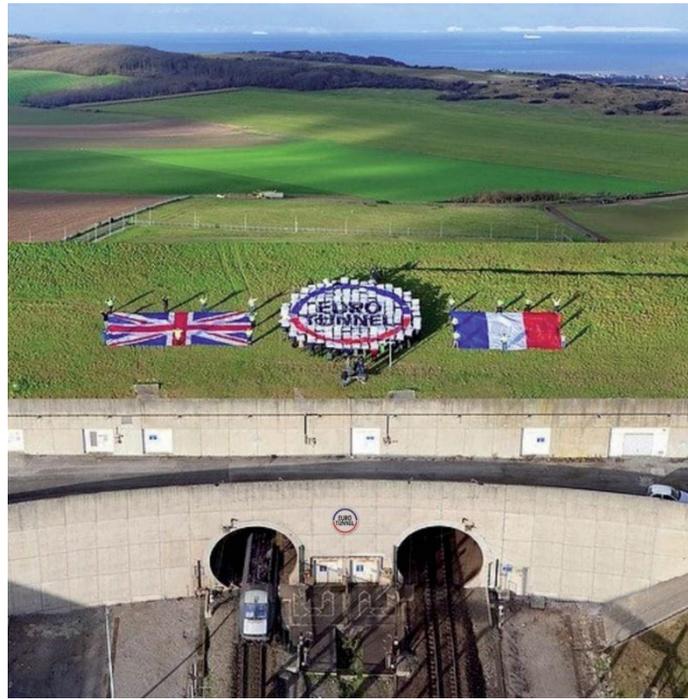

**Fig. 4  Eurotunnel.  A possible candidate for implementing zero energy travel. The tunnel entrances can be sealed to provide low pressure environment inside the tunnels. (Groupe Eurotunnel / 2010 Annual Review)**

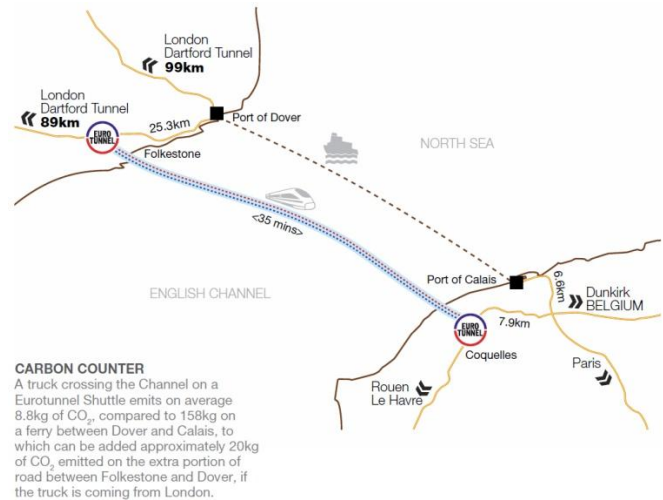

**Fig. 5  From Groupe Eurotunnel / 2010 Annual Review**